# A transient radio jet in an erupting dwarf nova

Elmar Körding[1*], Michael Rupen[2], Christian Knigge[1], Rob Fender[1], Vivek Dhawan[2], Matthew Templeton[3], Tom Muxlow[4]

[1] School of Physics and Astronomy, University of Southampton, Southampton SO17 1BJ, UK

[2] National Radio Astronomy Observatory, 10003 Lopezville Road, Socorrow NM 87801, USA

[3] American Association of Variable Star Observers, 49 Bay State Road, Cambridge, MA 02138, USA,

[4] University of Manchester, Jodrell Bank Observatory, Macclesfield SK11 9DL, UK

**Astrophysical jets seem to occur in nearly all types of accreting objects: from supermassive black holes to young stellar objects. Based on X-ray binaries, a unified scenario describing the disc/jet coupling has evolved and extended to many accreting objects. The only major exceptions are thought to be cataclysmic variables: Dwarf novae, weakly accreting white dwarfs, show similar outburst behaviour as X-ray binaries but no jet has yet been detected. Here we present radio observations of a dwarf nova in outburst showing variable flat-spectrum radio emission that is best explained as synchrotron emission originating in a transient jet. Both the inferred jet power and the relation to the outburst cycle are analogous to those seen in X-ray binaries, suggesting that the disc/jet coupling mechanism is ubiquitous.**

\* To whom correspondence should be addressed; Email: elmar@phys.soton.ac.uk

Jets launched by accreting objects seem to be an ubiquitous phenomenon, suggesting that accretion and jet launching may be intrinsically coupled. Jet emission from accreting white dwarfs (WDs) has been reported for supersoft sources (*1*, WDs with thermonuclear burning) and symbiotic stars (*2*, highly accreting binary systems). However, no jets have been found in cataclysmic variables (CVs), except perhaps after Nova eruptions (*3*). In fact, the lack of jet emission from dwarf novae (DNe), a class of weakly accreting non-magnetic CVs, has been used as a constraint for jet launching mechanisms of accreting objects (*4, 5*). Radio emission – often used as a tracer for a jet – has only sporadically been found for non-magnetic DNe (*6, 7*), and the radio detections are usually not reproducible (*8*). It has thus been suggested that the radio emission is correlated with the optical outburst (*9*). X-ray binaries (XRBs), which do show jets, share many properties with DNe: The triggering of an outburst as well as the subsequent evolution of the accretion disc (e.g., a truncated disc) are thought to be similar (*10*).

XRBs can be well studied throughout a full outburst cycle because the timescale from quiescence to the peak of the outburst and back ranges from weeks to months (*11*). The accretor in XRBs may be either a black hole or a neutron star. One of the main results of the study of black hole XRB is the establishment of accretion states (*12*) through which a source moves in a predefined order (*13*) and their associated jet properties. These states can be well separated on a hardness-intensity diagram (HID, *11*). At the beginning of the outburst, the source shows a hard X-ray spectrum and usually shows radio emission originating from a jet (*14*, the hard state, zone A in the left panel of figure 1). The source brightens while staying in the hard state until it makes a transition to the soft state characterized by a soft X-ray spectrum. This transition is typically accompanied by a bright radio flare once the source crosses the jet line, after this the core radio emission is quenched in the soft state (*13*). During the decay of the outburst the source moves back to the hard state – albeit at a lower luminosity than the hard to soft transition (*13*). While the nomenclature of neutron star XRB states is different, one can map the neutron star states onto the black hole equivalents (*15*). This can be visualized in an HID, where they follow basically the same pattern (*16*) as shown in the middle panel of figure 1. The main difference with respect to their radio emission is that the radio emission is only suppressed by a factor ~10 when the source is in the analogue state to the soft state (*17*). The different behaviour may be due to the existence of a boundary layer in neutron star XRBs, which does not exist in the black hole case.

The analogy between XRBs and DNe can be visualized by constructing a disc-fraction luminosity diagram (*18*) of a DN (figure 1 right panel), which is a generalization of the HIDs used for XRBs. We note that for a DN the inner region of the accretion flow is truncated by the stellar surface and its boundary layer. The boundary layer is thought to be the origin of the X-ray and extreme UV emission. The disc-fraction plotted in figure 1 describes the optical depth of the accretion flow for UV emission. The described analogy between XRBs and CVs suggests that radio emission from a DNe should be most prominent during the initial rise (zone A of figure 1) and during the subsequent state transition to the soft state. However, the timescale of this rise is usually of the order of 24h, making this phenomenon hard to catch.

In order to observe a DN in the radio band during the rise of an outburst, the American Association of Variable Star Observers (AAVSO) monitored a sample of ten DNe on our behalf. On April 24$^{th}$ 2007, we received notice from the AAVSO that the prototypical DN SS Cyg had brightened to a magnitude of 11.3 in the V-band, indicating the onset of an outburst. We subsequently triggered VLA observations at 8.6 GHz, which started approximately 10h after the initial optical observation. A typical observation (phase referenced to BL Lac) lasted for 2 hours and had a noise of 20 μJy/beam.

We detected SS Cyg at 8.5 GHz during this long optical outburst: slightly after the beginning of the outburst we detect a fast rise of the radio flux to 1.1 mJy that immediately declines again to a flux of ~ 0.3 mJy. This flux declines further with time, albeit more slowly than the optical emission (see figure 2). During the 1.1 mJy 'flare' we find upper limits for the linear polarization of 3.2 ± 2.7 % and for the circular polarization of -3.2 ± 2.7 %. During the decline of the outburst we also observed the source twice at 4.9 GHz, in addition to the 8.5 GHz observations. Both observations indicate that the source had a slightly inverted spectrum with an average spectral index of α = 0.3 ± 0.2 (flux $S_\nu \sim \nu^\alpha$).

We also detected SS Cyg with the 'Multi-Element Radio-Linked Interferometer Network' (MERLIN) at 1.66 GHz as a point source with 0.79 ± 0.10 mJy, 13 hours after the detection of the radio 'flare' with the VLA. These observations at higher angular resolution than the VLA indicate that the source of the radio emission is smaller than the beam size of ~0.2 arcsec. The position of the radio emission as measured by MERLIN is in agreement with the VLA position and coincides with the optical position of SS Cyg (nominal offset 27 mas when including proper motion). We did not detect any proper motion of the radio source associated with SS Cyg in our VLA images (average VLA beam size 11 x 7 arcsec, nominal offsets < 2 arcsec).

The only possibilities for radio emission from a CV are optically thick or thin thermal emission, synchrotron emission or coherent emission processes. From the angular resolution and flux of the MERLIN detection, we find a brightness temperature of at least 11000 K. The radio spectrum of optically thick thermal emission with such high temperatures has a spectral index of ~2, which would predict an 8.5 GHz VLA flux density at the time of the MERLIN observation of > 20 mJy in contrast to the observed light-curve. Optically thin thermal free-free emission could produce the measured spectrum. As the radio light-curve does not directly follow the bolometric luminosity of the CV, it is unlikely that the emitting gas cloud is detached from the accreting system and only reprocesses the energy emitted from the CV. Thus, the emitting gas is likely to originate from the CV. In case of a uncollimated outflow (a wind) we obtain an upper limit to the 8.6 GHz flux from optically thin thermal emission of $10^{-3}$ mJy (*19*) by assuming that this wind carries all of the accreted material (~$10^{-8}$ Msol/yr) into the emitting gas cloud. This is far below the measured value. If one collimates the outflow, i.e. one has a jet, one obtains higher fluxes. However, as the jet would have to carry nearly all accreted material and be extremely highly collimated (opening angle < 0.2°) to obtain the measured flux, we consider this an unlikely explanation.

Coherent emission is seen only in line emission (e.g., masers) and very steep-spectrum continuum emission, as for example observed in pulsars or flare stars (*20*). This is inconsistent with the measured flat spectrum. For gyro-synchrotron emission one would expect a high degree of circular polarization and our non-detection thus argues against this emission process. Additionally, the magnetic field of SS Cyg is thought to be fairly low (*21*) and should not play a dominant role in the emission mechanism: In order for the brightness temperature of the observed emission during the radio plateau not to exceed the Compton limit of ~ $10^{12}$ K, the size of the emission region has to be larger than ~60,000 km, more than 10 times the size of the central WD. The magnetosphere of SS Cyg in outburst is expected to be smaller than a few dwarf radii (*21*) – if existent at all. Thus, it is unlikely that the radio emission originates from any magnetic accretion processes near the WD. Synchrotron emission is known to produce a very high surface brightness, up to $10^{12}$ K, and may have spectral indices from -1.5 to 2.5 depending on optical depth. Thus, it is the best possibility for the observed emission.

As we have not resolved the emission region with our observations, it is hard to assess its geometry. The best known geometries are expanding shells and jets. Shells are commonly seen in explosive phenomena like Novae (*22*), but have not been seen in a normal DN outburst. Any synchrotron emission from a transient shell or jet ejection has a steep spectrum during the decline, while we observe a flat spectrum during the decay. In a jet scenario the observed behaviour can be created by having a compact radio jet during the initial rise of the outburst, then a transient ejection followed by a restarting compact jet. This is exactly what was suggested by the analogy between CVs and XRBs (figure 1), which are known to be jet emitters. SS Cyg does show radio emission during its soft state; this behaviour corresponds more closely to that observed in neutron stars, possibly because both neutron stars and WDs accrete onto a stellar surface and hence form a boundary layer.

For jet-emitting XRBs and active galactic nuclei (AGN) the radio emission in the hard state correlates well with the power liberated in the accretion flow (*23*). Taking into account that, for a given accretion rate, the accretion flow onto a WD liberates roughly 500 times less power than accretion onto a neutron star, we find $F_{8.6GHZ}$ = 0.44 mJy (Mdot/ $10^{-8}$ Msol yr$^{-1}$)$^{1.4}$ if the system is located at the HST-parallax distance of 166 pc (*24*). Typical DNe outbursts reach accretion rates of $\sim 10^{-8}$ Msol yr$^{-1}$, but the HST distance and peak brightness of SS Cyg would imply an anomalously high accretion rate of $\sim 10^{-7}$ Msol yr$^{-1}$ for this system in outburst (*25*). A smaller distance of $\sim$80 pc would be required to bring SS Cyg's peak accretion rate in line with typical values for DNe (*25*). Using this distance and an accretion rate of $\sim 5 \cdot 10^{-9}$ Msol yr$^{-1}$ during the rise one obtains a flux of $\sim$0.7 mJy, in agreement with the observations (the uncertainty of the correlation is a factor $\sim$2). However, if the HST distance is correct and the accretion rate is indeed as high as $\sim 10^{-7}$ Msol yr$^{-1}$, we have to assume an accretion rate during the rise of $\sim 5 \cdot 10^{-8}$ Msol yr$^{-1}$. The predicted radio flux is then $\sim$4 mJy, which is still roughly consistent with the measured values. In this case the jet emission of SS Cyg, especially during the optical plateau, may correspond to the highest accreting states in XRBs and not to the hard state. For such high accretion rates jet launching for CVs has been suggested by (*5*).

The similarities in the radio luminosity as well as in the relation of the radio emission to the accretion 'states' suggest that we did observe a jet from a non-magnetic DN. The detection of a jet in a DN with similar disc/jet coupling as seen in XRBs suggests that there may be common jet launching mechanism in CVs and XRBs. The radii of WDs lie roughly half-way in log-space between those seen in young stellar objects (YSOs) and those of black holes, so WDs connect YSOs to XRBs. With the exception of source classes similar to soft state XRBs, all accretion powered jet emitting sources from YSO (*26*), via AGN (*27*) to gamma ray bursts (*28*) seem to have a jet launching efficiency, i.e. the ratio of the jet power to the power liberated in the accretion flow, of $\sim$10 %. This suggests that there may be a common disc/jet coupling in all accreting objects from YSOs to gamma ray bursts.

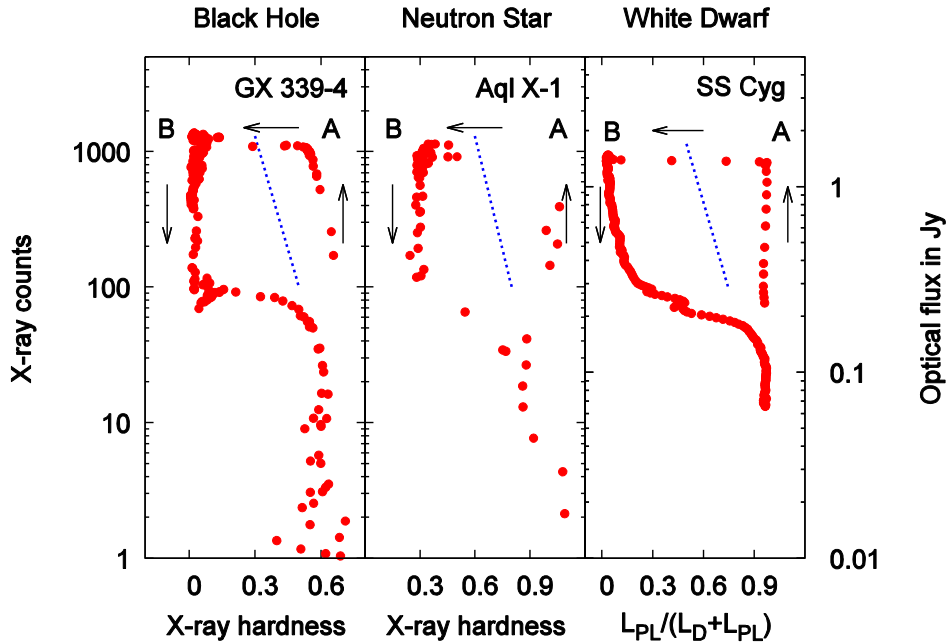

Figure 1: Hardness intensity diagrams (HID) for a black hole, a neutron star and the dwarf nova SS Cyg. The arrows indicate the temporal evolution of an outburst. The dotted line indicates the "jet line" observed in black hole and neutron star XRBs: on its right side one generally observes a compact jet; the crossing of this line usually coincides with a radio flare. For SS Cyg we show a disc-fraction luminosity diagram. Here, we plot optical flux against the power-law fraction measuring the prominence of the "power law component" in the hard X-ray emission in relation to the boundary layer/accretion disk luminosity. This power-law fraction has similar properties to the X-ray hardness used for XRBs. The diagram is based on data from (*29, 16*) and we use their conversion factors from extreme ultra-violet counts to disc/boundary layer luminosity $L_D$. The X-ray luminosity $L_{PL}$ for SS Cyg is for the 3-18 keV energy range. For the other objects the hardness ratio is defined as the ratio of the counts in the 6.3-10.5 keV range to 3.8 – 6.3 keV range, and the X-ray counts represent the 3.8 – 21.2 keV counts of the Rossi X-ray Timing Explorer.

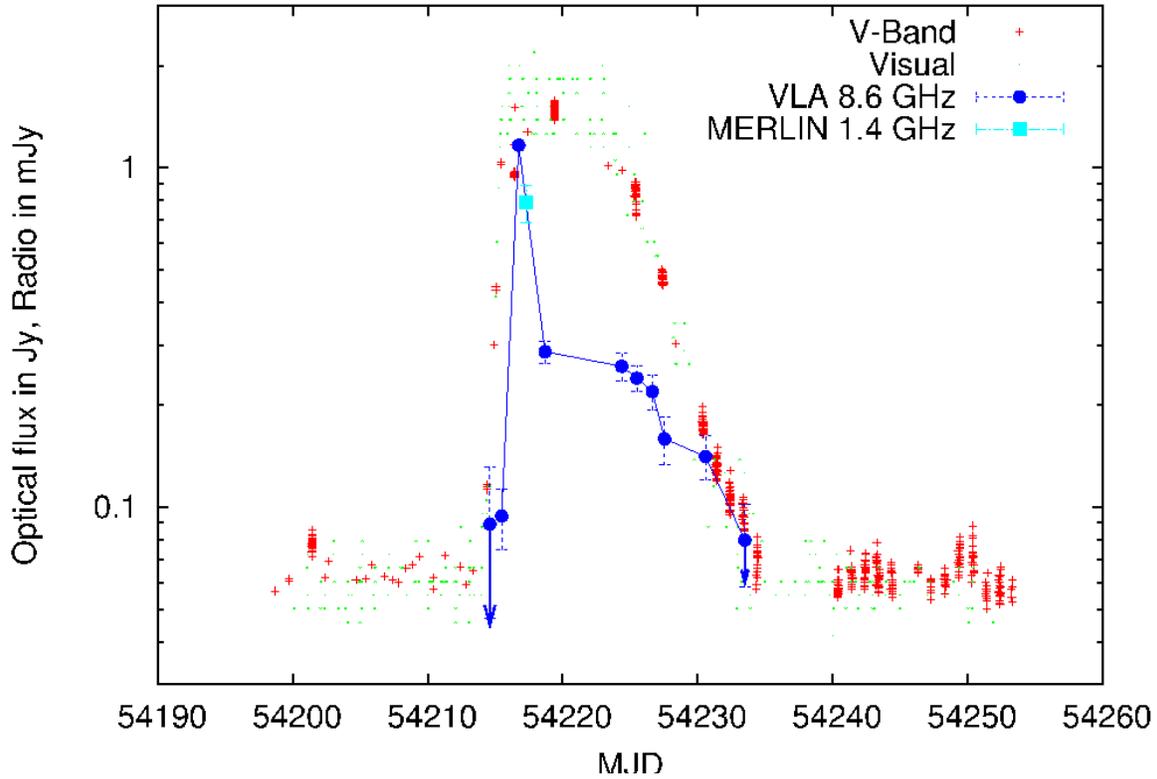

Figure 2: Radio and optical light-curve of SS Cyg. While the first observation does not detect SS Cyg significantly, the source is detected in the 8 following observations. After the first significant detection of 0.09 mJy, the source brightened to 1.1 mJy within 1.3 days and declined again to 0.29 mJy within 2 days. From there it declined more slowly than the optical light-curve until it was no longer detected in the last epoch (upper limit 0.08 mJy; rms 17 μJy/beam).

30. We acknowledge all variable star observers (AAVSO and others) for their monitoring of DNe. We are grateful to the fast scheduling by the scheduling officers of the VLA and MERLIN. EK acknowledges support from a Marie Curie IEF. The National Radio Astronomy Observatory is a facility of the National Science Foundation operated under cooperative agreement by Associated Universities, Inc. MERLIN is a National Facility operated by the University of Manchester at Jodrell Bank Observatory on behalf of STFC.